\documentclass[aps,prl,twocolumn,showpacs,preprintnumbers,amsmath,amssymb,
superscriptaddress]{revtex4}

\usepackage{graphicx}
\usepackage{dcolumn}
\usepackage{bm}
\begin{document}
\title{Exchange coupling in Eu monochalcogenides from first principles}
\author{J. Kune\v{s}}
\affiliation{Department of Physics, University of California, One Shields Avenue,
Davis CA 95616, USA}
\email{kunes@fzu.cz}
\affiliation{Institute of Physics,
Academy of Sciences of the Czech Republic, Cukrovarnick\'a 10,
162 53 Praha 6, Czech Republic}
\author{Wei Ku}
\altaffiliation[current address:\quad]{Department of
Physics,Brookhaven National Laboratory, managed by BSA for U.S. DOE under
Contract No. DE-AC02-98 CH 10886.}
\affiliation{Department of Physics, University of California, One Shields Avenue,
Davis CA 95616, USA}
\author{W.\,E. Pickett}
\affiliation{Department of Physics, University of California, One Shields Avenue,
Davis CA 95616, USA}
\date{\today}

\begin{abstract}
Using a density functional method with explicit account for strong Coulomb repulsion within 
the $4f$ shell, we calculate effective exchange parameters and the corresponding
ordering temperatures of the (ferro)magnetic insulating Eu
monochalcogenides (EuX; X=O,S,Se,Te) at ambient and elevated pressure conditions.
Our results provide quantitative account of the many-fold
increase of the Curie temperatures with applied pressure
and reproduce well the enhancement of the tendency toward ferromagnetic 
ordering across the series from telluride
to oxide, including the crossover from antiferromagnetic to 
ferromagnetic ordering under pressure in EuTe and EuSe. 
The first and second neighbor effective exchange are shown to
follow different functional dependencies. Finally, model calculations
indicate a significant contribution of virtual processes involving
the unoccupied $f$ states to the effective exchange.
\end{abstract}

\maketitle

Ferromagnetic semiconductors have become object of a great technological interest with
the appearance of spintronics because they can provide a spin-dependent tunneling 
barrier. Especially challenging is to
achieve a sizable ordered moment at room temperature, which is crucial for a large scale
application of the technology. The ability to calculate the ordering temperature and
understand the exchange mechanisms on the material specific level is of particular 
importance.
There is currently an intense effort to locate such materials within the dilute 
magnetic semiconductors,
where the magnetic moment is carried by impurities in an otherwise non-magnetic
system, but few candidates have been found. An attractive alternative is
FM insulators. Ferromagnetism is rare in stoichiometric materials without charge carriers.
Europium monochalcogenides (EuO, EuS, EuSe, EuTe) belong to this small group of
ferromagnetic insulators \cite{mau86}.

Crystallizing in the rock-salt structure, the first two members of the group order
ferromagnetically at 69.2 K and 16.6 K respectively \cite{pas76}, while EuTe, a type II antiferromagnet,
becomes ferromagnetic only at elevated pressure \cite{ish97}. EuSe is at the border line between
ferromagnetic and antiferromagnetic order with ferromagnetism stabilized by a moderate pressure 
of 0.5 GPa \cite{fuj82}. Application of pressure strongly enhances 
Curie temperatures of all these materials. 
The Eu$^{2+}$ valency results in half filling of the Eu $4f$ shell with $^8S$ configuration
of the groundstate multiplet.
Due to the localized nature of the moment-carrying $f$ orbitals, inter-site exchange 
interactions can only be mediated by the valence 
and conduction electrons. The intra-atomic $f-d$ and $f-s$ exchange, which
is the leading $f$-valence interaction, gives rise to temperature dependent features
(red shift effect) in the valence electron spectrum which are well captured 
by the ferromagnetic Kondo lattice model \cite{sch01,mul02}. It is due to the insulating
groundstate that this interaction alone cannot give rise to an effective coupling 
between the local moments.
To do so excitations across the gap and/or mixing of the $f$ and valence/conduction bands has to be 
taken into account. The relevant exchange processes have been discussed previously on a 
qualitative level \cite{kas70,saw76,lee84}. While a significant amount of {\it ab initio}
calculations of the Curie temperature in metallic systems has been done (e.g. Ref. \onlinecite{paj01} and
references therein) attempts 
to address the Curie temperature and coupling mechanisms of ferromagnetic insulators on a 
first principles level are
rare and become quite involved \cite{wei02,fel95}. 

The $f$ states in rare earths such as Eu pose an extra challenge in obtaining a quantitative microscopic
theory. {\it Ab initio} electronic structure methods
based on density functional theory (DFT) \cite{hoh64} and the
standard semi-local approximations \cite{koh65,per92} have notorious problems in dealing with 
the strong correlations within the $4f$ shell. In particular these approximations often 
result in incorrect filling of the $4f$ states. Early bandstructure calculation of europium 
chalcogenides by Cho \cite{cho70} using empirical potential did not address the 
inter-site exchange coupling. 
In this work we use the LDA+U method \cite{ani93},
which provides correct filling of the $4f$ states while allowing for mixing with the rest
of the band (unlike the frequently used open-core treatment). Just as importantly, positions of the resulting 
occupied and unoccupied $f$ bands are realistic.
The half filling of the $4f$ shell in Eu$^{2+}$ removes additional problems associated with orbital
degrees of freedom.

The calculations reported here were performed using the Wien2k\cite{wien2k} implementation of the 
full-potential linearized augmented-plane-waves (FLAPW) method with the rotationally invariant 
LDA+U functional and double-counting scheme of Ref. \onlinecite{ani93}. The size of
APW+lo basis was determined by the cut-off $R_{mt}K_{max}$=8 corresponding to approximately
100 basis functions per atom. Approximately 30 irreducible
k-points (depending on the magnetic structure) out of the 250-k-point regular grid were used in 
the Brillouin zone integrations.
The calculations were performed for lattice constants spanning the experimental range of stability
of the rock salt crystal structure.
\begin{figure}
\includegraphics[height=8.5cm,angle=270]{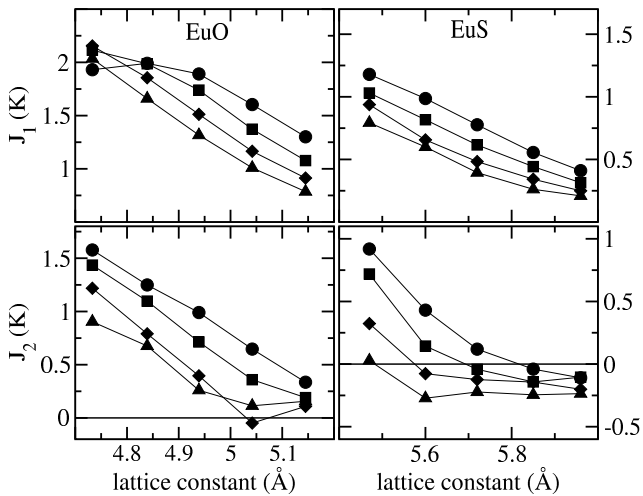}
\caption{\label{fig:jos} The nearest-neighbor $J_1$ and next-nearest-neighbor $J_2$ effective
exchange parameters in EuO and EuS as functions of the lattice constant calculated for different values of U
(circle -- 6 eV, square -- 7 eV, diamond -- 8 eV, triangle -- 9 eV, the lines serve as guides for 
eye). The deviation for the linear dependence of $J_1$ at high pressures for EuO coincides with
the onset of metallic behavior due to overlap of the $f$ and valence band.}
\end{figure}
The groundstate energies of three different magnetic structures:
(i) ferromagnetic (F), 
(ii) type II antiferromagnetic and
(iii)  antiferromagnetic with propagation vector (0,0,$\tfrac{2\pi}{a}$),
were calculated self-consistently and mapped on Heisenberg Hamiltonian 
\begin{equation}
H=-\sum_{i,j} J_{ij}\mathbf{S}_i\cdot\mathbf{S}_j 
\end{equation}
with nearest-neighbor $J_1$ and next-nearest-neighbor $J_2$ interaction, which is 
known experimentally to provide a good description of magnetic behavior
of the materials in question.  The classical energies corresponding to spin 
configurations (i) to (iii) are $-(12J_1+6J_2)S^2$,
$6J_2S^2$ and $(4J_1-6J_2)S^2$ respectively. The ferromagnetic energies were calculated
with the lattices of both antiferromagnetic structures and the corresponding
energy differences were taken.
The assumption of $J_1-J_2$ model was checked
for EuS at ambient pressure by 
spin-spiral calculation similar to those of Ref. \onlinecite{kun04}, 
which yielded dispersion consistent with $J_1-J_2$ model. 
The calculated exchange constants as function of lattice parameter for values of U from 6 eV to 9 eV are shown
in Fig. \ref{fig:jos} and Fig. \ref{fig:jsete}. 
\begin{figure}
\includegraphics[height=8.5cm,angle=270]{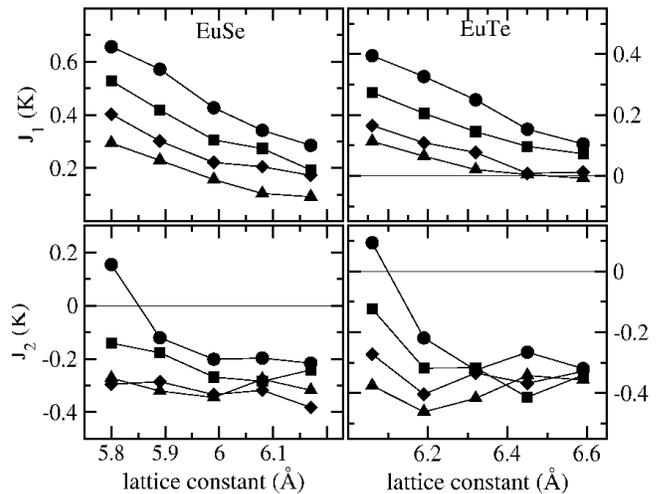}
\caption{\label{fig:jsete} The same as in Fig. \ref{fig:jos} for EuSe and EuTe.}
\end{figure}

\begin{figure}
\includegraphics[height=8.5cm,angle=270]{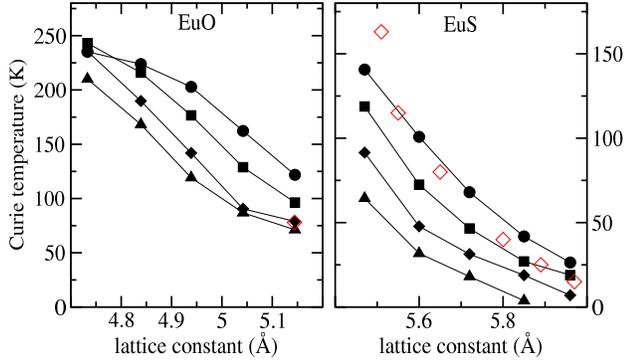}
\caption{\label{fig:tos} The magnetic ordering temperatures of EuO and EuS calculated for
different values of U (see caption of Fig. \ref{fig:jos}). The experimental values
for EuO \cite{zin76} and EuS \cite{gon98} are marked with open diamonds.}
\end{figure}

\begin{figure}
\includegraphics[height=8.5cm,angle=270]{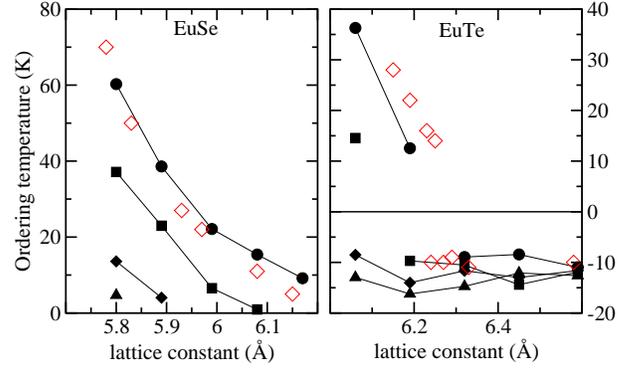}
\caption{\label{fig:tsete} The magnetic ordering temperature of EuSe and EuTe calculated for
different values of U (see caption of Fig. \ref{fig:jos}). The negative values for EuTe correspond
to N\'eel temperature and antiferromagnetic groundstate. The experimental values marked with
open diamonds are taken from Ref. \cite{gon97}.}
\end{figure}

From the parameters of the Heisenberg Hamiltonian ordering temperatures were calculated
using result of the Tyablikov decoupling method \cite{tya67,tah62}
\begin{equation}
\label{eq:tyab}
(k_BT_{C})^{-1}=\frac{2S(S+1)}{3}\frac{1}{N}\sum_{\mathbf q}
\bigl[J({\mathbf 0})-J({\mathbf q})\bigr]^{-1},
\end{equation}
where $J({\mathbf q})$ stands for the lattice Fourier transform of the effective exchange parameter. 
The alignment in the groundstate is determined by the sign of $J_1+J_2$ 
(positive--ferromagnetic, negative--antiferromagnetic). For EuTe we have calculated
the N\'eel temperature as well using generalized equation (\ref{eq:tyab}) \cite{tur03}.
The ordering temperatures as functions of lattice constant are shown in Fig. \ref{fig:tos} and
Fig. \ref{fig:tsete} and compared to the experimental data of Goncharenko and Mirebeau \cite{gon97,gon98}
(see also \cite{fuj82,ish97}).
The estimated numerical accuracy is about 10 K.
The trend of weakening ferromagnetism in favor of antiferromagnetism 
when going from oxide to telluride is well reproduced and is common to all values of 
U. The effective exchange is rather sensitive to the value of U, yet the literature
value of 6 eV to 7 eV \cite{har95}
gives the best agreement throughout the 
series. Both the significant increase of the Curie temperature with pressure 
and almost constant behavior of the N\'eel temperature of EuTe at lower pressures
observed experimentally are well captured by the calculations

Based on the fact that in the type II antiferromagnetic
structure the first neighbor exchange is frustrated and thus the mean-field N\'eel temperature 
is proportional to $J_2$, Goncharenko and
Mirebeu concluded that $J_2$ is pressure independent while $J_1$ exhibits a non-linear
increase with the applied pressure. In their scenario the transition from low-pressure antiferromagnetic
groundstate to high-pressure ferromagnetic groundstate is solely due to the increase of $J_1$. 
Our calculations provide a different picture.
With the only exception being EuO at high pressure, we find more
or less linear dependence of $J_1$ on the lattice parameter. On the other hand $J_2$ exhibits quite
non-linear behavior, which in the case of EuTe translates to being almost constant at low pressures and
increasing rapidly at higher pressures, which significantly contributes to stabilization
of the ferromagnetic state.

\begin{figure}
\includegraphics[height=8.5cm,angle=270]{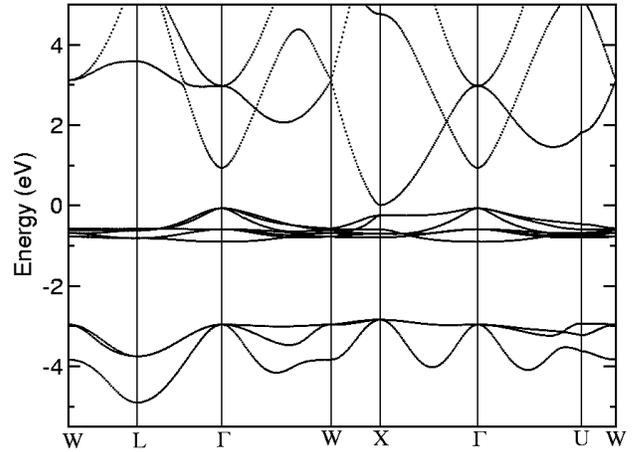}
\caption{\label{fig:band} Majority spin bandstructure of EuO obtained with U of 7 eV at ambient pressure.
The valence bands have dominant O $2p$ character while the conduction bands are mostly Eu $d$ and $s$
states.}
\end{figure}

The success of the LDA+U functional in describing the trend across the chalcogenide series as well
as capturing the pressure dependence of the ordering temperature indicates that the relevant coupling
mechanisms are well accounted for. While we can not identify directly the leading coupling mechanisms we can
make several observations connected to the sensitivity of the results to the value of U. The value
of U affects inter-site coupling in two distinct ways: (i) through the position of the occupied
$f$ bands within the semiconducting gap, which is decisive for hybridization with
the valence and conduction bands, (ii) through the splitting between the occupied and unoccupied
$f$ bands. In order to stress the importance of including hybridization effects in the $f$ bands
we show the spin-majority bandstructure of EuO in Fig. \ref{fig:band}. The $f$ band exhibits
appreciable dispersion with a bandwidth of about 1 eV which is mostly due to mixing with O $2p$ 
valence bands. A similar picture is obtained for the other members of the series with the $f$ bands 
being localized deeper in the gap when going toward telluride.

\begin{figure}
\includegraphics[angle=270,width=8.5cm]{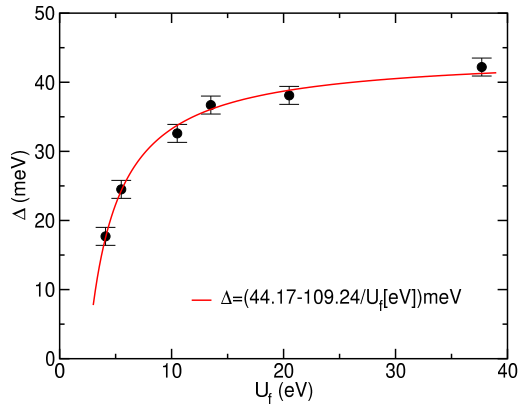}
\caption{\label{fig:model} Difference between the groundstate energies of type II antiferromagnetic and
ferromagnetic structures as a function of the splitting $U_f$ between the occupied and unoccupied
$f$ bands obtained with auxiliary potential on top of LDA+U with U of 7 eV. $U_f$ of 11 eV
corresponds to zero auxiliary potential.}
\end{figure}
The intra-atomic $f-s$ or $f-d$ exchange is determined solely by the spin density.
The coupling mechanisms involving only the intra-atomic exchange
are therefore insensitive to the value of U. 
The Bloembergen-Rowland (BR) coupling \cite{blo55}, inter-band analog of the Ruderman-Kittel-Kasuya-Yosida 
(RKKY) coupling well known in metals, is the leading mechanism of this type. A discussion
of BR coupling was given earlier by Lee and Liu \cite{lee84}, starting from diagonalized band Hamiltonian
and adding interband exchange term, and by Kasuya \cite{kas70} and Sawatzky {\it et al.} \cite{saw76},
starting from atomic orbitals with $f-d$ exchange and conduction-valence (Eu $d$ - X $p$) hybridization.
The sizable dependence of our results on the value
of U indicates that additional mechanisms are involved. 

We have performed a simple model
calculation to evaluate the role of unoccupied $f$ bands which participate, for example, in the superexchange
mechanism. By adding an orbital dependent auxiliary potential, which acts only on the minority-spin
$f$ orbitals, we control the 'site energy' of the unoccupied (minority-spin) $f$ states 
without effecting the occupied $f$'s. Obviously such a term does not enter the groundstate
energy directly, but only through the mixing of the minority-spin $f$ bands with the occupied bands.
Excitations from occupied to unoccupied bands correspond to hopping of an $f$ electron from one
Eu atom to another, resulting in superexchange interaction. By looking at the dependence of the
effective exchange parameters on the splitting of the occupied and  unoccupied $f$ bands controlled
by the auxiliary potential we can assess the relative importance 
of the superexchange and virtual processes involving change in the $4f$ occupation in general.
In Fig. \ref{fig:model} we show the energy difference between the ferromagnetic and type II antiferromagnetic
groundstates as a function of energy separation between the occupied and unoccupied $f$ bands.
Apparently the auxiliary potential has a sizable effect consistent with $1/U_f$ dependency of the superexchange
interaction. 

Now we summarize. Our calculations show that accounting for intra-atomic repulsion using the 
LDA+U method provides a reliable description of
effective exchange coupling in ferromagnetic insulators with localized moments. 
The trend favoring ferromagnetism for lighter chalcogenides as well as the under-pressure
antiferro-to-ferromagnetic transition in EuSe and EuTe are well captured.
The pressure dependences of the magnetic ordering temperatures, which correspond well
to the experimental observations, are connected to distinct under-pressure behavior
of the exchange parameters $J1$ and $J2$.
The exchange coupling is strongly 
effected even by the unoccupied $f$ bands establishing the important role of virtual hopping processes 
between the $f$ shells on neighboring atoms.

This work was supported by NSF Grant DMR-0114818 and by the SSAAP program (DE-FG03-03NA00071)
at University of California Davis.

\end{document}